\documentclass[a4paper]{spie}

\usepackage{amsmath,amsfonts,amssymb,graphicx}
\usepackage[colorlinks=true,allcolors=blue]{hyperref}
\usepackage{times,mathpazo}

\title{Routing thermal noise through quantum networks}

\newcommand{\rmd}{\mathrm{d}}

\author[a]{Andr\'e Xuereb}
\author[b]{Matteo Aquilina}
\author[c]{Shabir Barzanjeh}
\affil[a]{Department of Physics, University of Malta, Msida MSD\,2080, Malta}
\affil[b]{National Aerospace Centre, Luqa LQA\,9023, Malta}
\affil[c]{Institute of Science and Technology Austria, 3400 Klosterneuburg, Austria}
\authorinfo{Please send correspondence to A.X.\ or S.B.\\E-mail: \href{mailto:andre.xuereb@um.edu.mt}{andre.xuereb@um.edu.mt} (A.X.), \href{mailto:shabir.barzanjeh@ist.ac.at}{shabir.barzanjeh@ist.ac.at} (S.B.)}

\begin{document} 
\maketitle

\begin{abstract}
There is currently significant interest in operating devices in the quantum regime, where their behaviour cannot be explained through classical mechanics. Quantum states, including entangled states, are fragile and easily disturbed by excessive thermal noise. Here we address the question of whether it is possible to create non-reciprocal devices that encourage the flow of thermal noise towards or away from a particular quantum device in a network. Our work makes use of the cascaded systems formalism to answer this question in the affirmative, showing how a three-port device can be used as an effective thermal transistor, and illustrates how this formalism maps onto an experimentally-realisable optomechanical system. Our results pave the way to more resilient quantum devices and to the use of thermal noise as a resource.
\end{abstract}


\section*{COPYRIGHT NOTICE}
Andr\'e Xuereb, Matteo Aquilina, and Shabir Barzanjeh, ``Routing thermal noise through quantum networks,'' eds.\ David L. Andrews, Angus J. Bain, Jean-Michel Nunzi, and Andreas Ostendorf, Proc. SPIE \textbf{10672} Nanophotonics VII, 10672N (2018).

Copyright 2018 Society of Photo-Optical Instrumentation Engineers. One print or electronic copy may be made for personal use only. Systematic reproduction and distribution, duplication of any material in this paper for a fee or for commercial purposes, or modification of the content of the paper are prohibited.

\href{https://doi.org/10.1117/12.2309928}{https://doi.org/10.1117/12.2309928}

\section{INTRODUCTION}
\begin{figure}[ht]
 \begin{center}
   \includegraphics[scale=0.4]{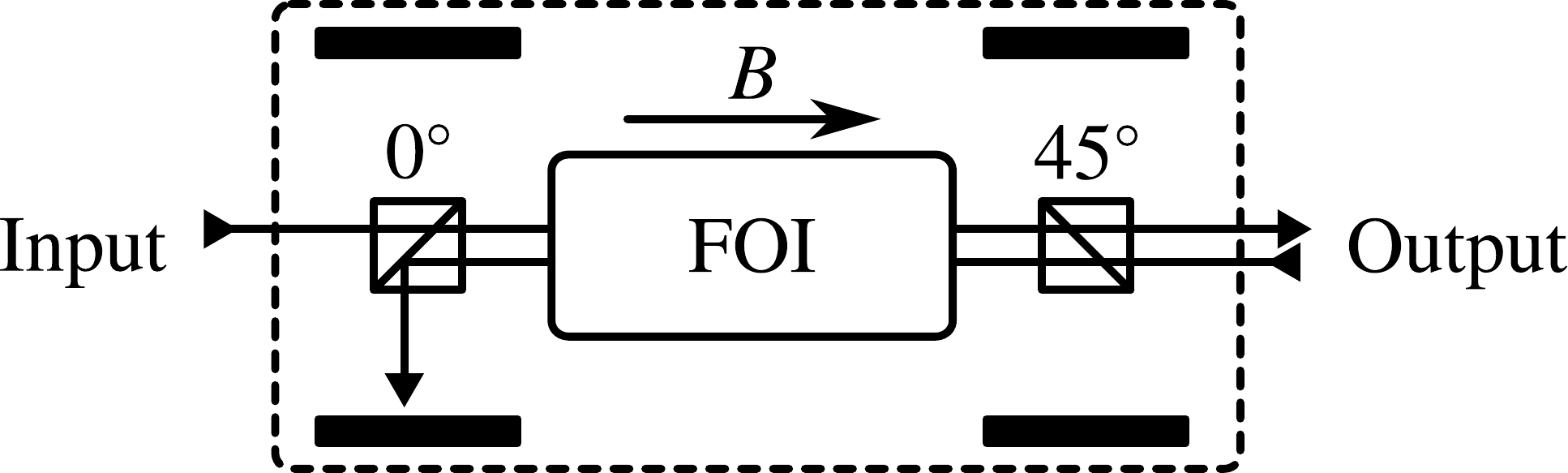}
 \end{center}
 \caption{\label{fig:FOI}Schematic of a Faraday optical isolator (FOI). A static magnetic field $B$ acts on a material to rotate the polarisation of light passing through it. We consider rotation by $45^\circ$ as a simple case. Two polarising beam-splitters, one at each end of the Faraday material and orientated at $45^\circ$ with respect to each other, complete the setup. Light entering the setup from the input port passes straight through, albeit suffering a change in polarisation; any back-reflection is diverted to a beam dump. Seen as a two-port device, as indicated by the dashed box, the FOI allows light to pass through in one direction but not in reverse.}
\end{figure}

Non-reciprocal devices for electromagnetic radiation are of significant practical use; much like the humble and ubiquitous diode in electronic circuits, the possibility to direct optical or microwave radiation in one direction, but not in reverse, is of importance in everything from telecommunications\cite{Kobayashi1980} to preventing feedback-induced instabilities in lasers\cite{Ohtsubo2013}. The paradigmatic example of a non-reciprocal device for light is a Faraday optical isolator (FOI), whose operation depends on the Faraday effect. The action of a magnetic field on specific materials causes the polarisation vector of light passing through the isolator to rotate in a specific direction. Due to the symmetry-breaking caused by the magnetic field, this rotation is not undone when light travels in the reverse direction. Upon interchanging the inputs and outputs of a FOI, one obtains qualitatively different behaviour (see Fig.~\ref{fig:FOI}).

This picture provides us with the foundation for a definition of non-reciprocal behaviour that is more general than symmetry breaking under time reversal. In dissipative systems, time-reversal symmetry is a weaker concept\cite{Lax1980}. Consider a dissipative system with multiple inputs and outputs; its dissipative nature implies that any transient input signal will cause a decaying output. Even if such a system were to be reciprocal, in the sense that the interchange of inputs and outputs yields identical physics, it could not be time-symmetric, since dissipation implies that any output signal will decay. Our investigation concentrates on dissipative systems, and we will correspondingly be concerning ourselves with systems whose behaviour is not identical when inputs and outputs are interchanged.

Whereas it is not our intention to review the significant process that has taken place in proposing and demonstrating non-reciprocal devices in recent years, it is useful to highlight a few specific approaches. Optomechanical systems\cite{Aspelmeyer2014} provide a veritable playground for investigating the interaction between light and motion at the nano- and micro-scales. This interaction was suggested as the basis for a non-reciprocal device by Hafezi and Rabl in 2012\cite{Hafezi2012}. Consider a toroidal optical micro-resonator supporting degenerate clockwise and counter-clockwise modes, coupled to a waveguide held close to it. Transmission along either direction in the waveguide is identical, due to the degeneracy of the modes in the resonator. One can break the symmetry by pumping one of these modes, e.g., the clockwise mode, thereby enhancing the optomechanical interaction between the mechanical breathing modes of the resonator and its optical modes. Under these conditions, transmission in one direction is preferred to that in the opposite direction; external pumping renders the system non-reciprocal. An alternative approach\cite{Peano2015} makes use of the optomechanical interaction in a specially-designed crystal structure. Tailored input optical fields, carefully chosen to pump each cell in the crystal with a specific phase, induce phases when photons or phonons hop from one cell to the next. This can be rephrased as an effective pseudo-magnetic field acting on the photons or phonons, thereby inducing non-reciprocal behaviour in their motion along the crystal. Further studies predicted topologically-protected edge states in mechanical systems that can even be used to build directional acoustic amplifiers\cite{Peano2016}.

The mechanisms described so far do not require dissipation to work, and can be described to an extent in a fully unitary picture. In this paper we will be concerned with non-unitary devices connected to heat baths. Under the guise of reservoir engineering\cite{Metelmann2015,Malz2018} parts of our work have been discussed previously. Such techniques have recently been exploited in optomechanical experiments to build circulators and non-reciprocal devices for microwave\cite{Bernier2017,Barzanjeh2017} and optical\cite{Ruesink2018} signals.

In this paper, however, we will take a different point of view in two essential ways\cite{Barzanjeh2018}. First, we will describe the system using the cascaded-systems formalism; despite this being textbook material\cite{Gardiner2004} we will briefly review its key points. Second, we will concentrate not on input and output \emph{signals}, but on the flow of thermal noise through a non-reciprocal network of quantum devices. Throughout, our focus will be on optomechanical devices as a platform on which to realise our proposal.

\section{THE CASCADED QUANTUM SYSTEMS FORMALISM}
Our basic building block is a network composed of two open quantum systems, which we label 1 and 2. We demand, by construction, that the output of system 1 forms the input of system 2, but not vice versa. Suppose that the two systems are single-mode bosonic fields, with which we associate annihilation operators $\hat{c}_1$ and $\hat{c}_2$. With each system $i=1,2$ we also associate an input field $\hat{b}_{\text{in},i}$, an output field $\hat{b}_{\text{out},i}$, and a decay rate $\gamma_i>0$ which sets the coupling rate between system $i$ and its input and output fields. As is well-known\cite{Gardiner2004}, the input--output formalism yields
\begin{equation}
\hat{b}_{\text{out},i}=\hat{b}_{\text{in},i}+\sqrt{\gamma_i}\hat{c}_i,
\end{equation}
which is to be interpreted in the following as a Heisenberg-picture equation with all the operators evaluated at some time $t$. To induce cascaded dynamics, we require further that $\hat{b}_{\text{in},2}(t)=\hat{b}_{\text{out},1}(t-\tau)$. The quantity $\tau\geq0$ denotes the time delay incurred between the two system, but we can formally set it to zero by shifting the time coordinate for system 2. The Langevin equation governing the evolution of any operator $\hat{a}$ of the compound system is\cite{Gardiner2004}
\begin{multline}
\label{eq:LE}
\tfrac{\rmd}{\rmd t}\hat{a}=-\tfrac{\imath}{\hbar}\bigl[\hat{a},\hat{H}_\text{sys}\bigr]-\bigl[\hat{a},\hat{c}_1^\dagger\bigr]\bigl(\tfrac{\gamma_1}{2}\hat{c}_1+\sqrt{\gamma_1}\hat{b}_{\text{in},1}\bigr)+\bigl(\tfrac{\gamma_1}{2}\hat{c}_1^\dagger+\sqrt{\gamma_1}\hat{b}_{\text{in},1}^\dagger\bigr)\bigl[\hat{a},\hat{c}_1\bigr]\\
-\bigl[\hat{a},\hat{c}_2^\dagger\bigr]\bigl(\tfrac{\gamma_2}{2}\hat{c}_2+\sqrt{\gamma_2}\hat{b}_{\text{in},1}\bigr)+\bigl(\tfrac{\gamma_2}{2}\hat{c}_2^\dagger+\sqrt{\gamma_2}\hat{b}_{\text{in},1}^\dagger\bigr)\bigl[\hat{a},\hat{c}_2\bigr]\\
-\bigl[\hat{a},\hat{c}_2^\dagger\bigr]\sqrt{\gamma_1\gamma_2}\hat{c}_1+\sqrt{\gamma_1\gamma_2}\hat{c}_1^\dagger\bigl[\hat{a},\hat{c}_2\bigr],
\end{multline}
where $\hat{H}_\text{sys}$ is the Hamiltonian that governs the evolution of systems $1$ and $2$; $\bigl[\hat{a},\hat{b}\bigr]=\hat{a}\hat{b}-\hat{b}\hat{a}$ is the commutator. The identification of the output of system 1 with the input of system 2 has effectively reduced the number of input and output ports of the network, which now has one global input, $\hat{b}_{\text{in},1}$, and one global output, $\hat{b}_{\text{out},2}$.

The Langevin equation~(\ref{eq:LE}) is equivalent to a master equation for the density matrix $\rho$ of the system in standard (Lindblad) form:
\begin{equation}
\label{eq:StdME}
\tfrac{\rmd}{\rmd t}\rho=-\tfrac{\imath}{\hbar}\bigl[\hat{H}_\text{eff},\rho\bigr]+\mathcal{D}_{\bar{N}_3,\kappa_3,\hat{c}_3}[\rho],
\end{equation}
where $\mathcal{D}_{\bar{N},\kappa,\hat{a}}[\rho]$ is the Liouvillian corresponding to a bosonic heat bath with average occupancy $\bar{N}$ that is coupled through system operator $\hat{c}$ with a rate $\kappa$:
\begin{equation}
\mathcal{D}_{\bar{N},\kappa,\hat{a}}[\rho]:=(\bar{N}+1)\kappa\bigl(\hat{a}\rho\hat{a}^\dagger-\tfrac{1}{2}\bigl\{\rho,\hat{a}^\dagger\hat{a}\bigr\}\bigr)+\bar{N}\kappa\bigl(\hat{a}^\dagger\rho\hat{a}-\tfrac{1}{2}\bigl\{\rho,\hat{a}\hat{a}^\dagger\bigr\}\bigr),
\end{equation}
with $\bigl\{\hat{a},\hat{b}\bigr\}=\hat{a}\hat{b}+\hat{b}\hat{a}$ being the anticommutator. In Eq.~(\ref{eq:StdME}) we introduced an effective Hamiltonian
\begin{equation}
\hat{H}_\text{eff}:=\hat{H}_\text{sys}+\tfrac{\imath\hbar}{2}\sqrt{\gamma_1\gamma_2}\bigl(\hat{c}_1^\dagger\hat{c}_2-\hat{c}_1\hat{c}_2^\dagger\bigr),
\end{equation}
a collective coupling rate $\kappa_3:=\gamma_1+\gamma_2$, and a collective bosonic annihilation operator
\begin{equation}
\hat{c}_3:=\tfrac{1}{\sqrt{\kappa_3}}\bigl(\sqrt{\gamma_1}\hat{c}_1+\sqrt{\gamma_2}\hat{c}_2\bigr),
\end{equation}
which satisfies $\bigl[\hat{c}_3,\hat{c}_3^\dagger\bigr]=1$. The physical content of master equation~(\ref{eq:StdME}) is that our cascaded quantum system is fully equivalent to two bosonic modes that are coupled to each other by means of the second (``hopping'') term in $\hat{H}_\text{eff}$, and which are coupled to a common bath by means of the collective damping operator $\hat{c}_3$. The non-reciprocal behaviour arises from the interference that is set up between these two channels, since excitations can hop between the two systems either through the former (unitary dynamics), or through the latter (non-unitary dynamics). The sign change between the hopping term and the operator $\hat{c}_3$ is the mathematical basis for the constructive (destructive) interference in the direction $1\to2$ ($2\to1$).

We will investigate a more complete situation, where each system is coupled independently to a heat bath as well as to the common heat bath described above. In order to make our comparison with the optomechanical situation in the next section more straightforward we introduce a phase $\phi$ to $\hat{c}_2$, and to account for imperfect non-reciprocity, we introduce a new coupling term to our Hamiltonian. Thus,
\begin{equation}
\hat{H}_\text{eff}:=\hat{H}_\text{sys}+\tfrac{\imath\hbar}{2}\sqrt{\gamma_1\gamma_2}\bigl(e^{\imath\phi}\hat{c}_1^\dagger\hat{c}_2-e^{-\imath\phi}\hat{c}_1\hat{c}_2^\dagger\bigr)+\hbar\bigl(F\hat{c}_1^\dagger\hat{c}_2+F^\ast\hat{c}_1\hat{c}_2^\dagger\bigr),
\end{equation}
where $F$ is an arbitrary complex number, and
\begin{equation}
\hat{c}_3:=\tfrac{1}{\sqrt{\kappa_3}}\bigl(\sqrt{\gamma_1}\hat{c}_1+\sqrt{\gamma_2}e^{\imath\phi}\hat{c}_2\bigr).
\end{equation}
Perfect non-reciprocity is restored when $F=0$. Finally, our full master equation reads
\begin{equation}
\tfrac{\rmd}{\rmd t}\rho=-\tfrac{\imath}{\hbar}\bigl[\hat{H}_\text{eff},\rho\bigr]+\sum_{i=1,2,3}\mathcal{D}_{\bar{N}_i,\kappa_i,\hat{c}_i}[\rho],
\end{equation}
where the Liouvillian term with $i=1,2$ corresponds to the heat bath for system $i$, with average occupancy $\bar{N}_i$, that is coupled to the network through the damping operator $\hat{c}_i$ with a coupling rate $\kappa_i$.

At this stage we need to specify $\hat{H}_\text{sys}$. We assume that the two systems are uncoupled bosonic modes, with free oscillation frequencies $\omega_i$ ($i=1,2$):
\begin{equation}
\hat{H}_\text{sys}=\hbar\omega_1\hat{c}_1^\dagger\hat{c}_1+\hbar\omega_2\hat{c}_2^\dagger\hat{c}_2.
\end{equation}
Starting from the Langevin equation~(\ref{eq:LE}) it is straightforward to show that
\begin{equation}
\tfrac{\rmd}{\rmd t}\hat{c}_1=-(\imath\omega_1+\tfrac{\gamma_1+\kappa_1}{2})\hat{c}_1-\imath F\hat{c}_2+\sqrt{\kappa_1}\hat{c}_{\text{in},1}+\sqrt{\gamma_1}\hat{c}_{\text{in},3},
\end{equation}
and
\begin{equation}
\tfrac{\rmd}{\rmd t}\hat{c}_2=-(\imath F^\ast+\sqrt{\gamma_1\gamma_2}e^{\imath\phi})\hat{c}_1+\sqrt{\kappa_2}\hat{c}_{\text{in},2}-(\imath\omega_2+\tfrac{\gamma_2+\kappa_2}{2})\hat{c}_2+\sqrt{\gamma_2}e^{\imath\phi}\hat{c}_{\text{in},3}.
\end{equation}
It is now easy to see that when $F=0$, system 2 is affected by system 1, but system 1 is entirely uncoupled from system 2. In these equations of motion, each input bosonic operator $\hat{c}_{\text{in},i}$ is associated with bath $i$ and has the following properties:
\begin{align}
\langle\hat{c}_{\text{in},i}(t)\rangle&=0,\\
\langle\hat{c}_{\text{in},i}^\dagger(t)\hat{c}_{\text{in},j}(t^\prime)\rangle&=\bar{N}_i\delta_{i,j}\delta(t-t^\prime),\ \text{and}\\
\langle\hat{c}_{\text{in},i}(t)\hat{c}_{\text{in},j}^\dagger(t^\prime)\rangle&=\bigl(\bar{N}_i+1)\delta_{i,j}\delta(t-t^\prime),
\end{align}
where $\delta_{i,j}$ is the Kronecker delta, $\delta(t-t^\prime)$ the Dirac delta function, and $i,j=1,2,3$.

In steady state, these equations can be Fourier-transformed from the time domain to the frequency domain. We can express the resulting equations compactly in matrix form:
\begin{equation}
-\imath\omega\begin{pmatrix}
\hat{c}_1 \\
\hat{c}_2
\end{pmatrix} = \begin{bmatrix}
-\imath\omega_1-\tfrac{\gamma_1+\kappa_1}{2} & -\imath F \\
-\imath F^\ast-\sqrt{\gamma_1\gamma_2}e^{\imath\phi} & -\imath\omega_2-\tfrac{\gamma_2+\kappa_2}{2}
\end{bmatrix}\begin{pmatrix}
\hat{c}_1 \\
\hat{c}_2
\end{pmatrix}
+\begin{pmatrix}
\sqrt{\kappa_1}\hat{c}_{\text{in},1} \\
\sqrt{\kappa_2}\hat{c}_{\text{in},2}
\end{pmatrix}+\begin{pmatrix}
\sqrt{\gamma_1} \\
\sqrt{\gamma_2}e^{\imath\phi} \\
\end{pmatrix}\hat{c}_\text{in,3}.
\end{equation}

In the next section we will develop an optomechanical system that realises this model.

\section{AN OPTOMECHANICAL SCENARIO}
We consider a network composed of two electromagnetic cavity modes that mutually interact with a mechanical oscillator via the standard optomechanical interaction. The Hamiltonian that generates the dynamics of this system is
\begin{equation}
\hat{H}=\hbar\omega_\text{m}\hat{b}^\dagger\hat{b}+\sum_{i=1,2}\hbar\bigl[\omega_i\hat{a}^\dagger_i\hat{a}_i+g_i\hat{a}^\dagger_i\hat{a}_i\bigl(\hat{b}+\hat{b}^\dagger\bigr)\bigr]
+\hbar J\bigl(\hat{a}_1\hat{a}_2^\dagger+\hat{a}_1^\dagger\hat{a}_2\bigr)+\sum_{i=1,2}\hbar\mathcal{E}_i\big(e^{-\imath\omega_{\text{d},i}t}\hat{a}_i+e^{\imath\omega_{\text{d},i}t}\hat{a}_i^\dagger\bigr),
\end{equation}
where $\hat{a}_i$ ($i=1,2$) is the bosonic annihilation operator that corresponds to electromagnetic mode $i$ whose frequency is $\omega_i$, and $\hat{b}$ is the annihilation operator corresponding to the mechanical oscillator with frequency $\omega_\text{m}$. Photons are allowed to hop directly between the electromagnetic modes; this process is governed by the coupling constant $J$, and each electromagnetic field mode is driven by means of a classical source with strength $\mathcal{E}_i$ and frequency $\omega_{\text{d},i}$. Finally, we describe the optomechanical interaction by means of the constant $g_i$, which shifts the position of the mechanical oscillatior by an amount proportional to the photon number of mode $i$. We can rewrite this equation in a frame rotating at the driving frequencies. Assuming that $\omega_{\text{d},1}=\omega_{\text{d},2}$, and defining $\Delta_i:=\omega_i-\omega_{\text{d},i}$, we obtain the time-independent Hamiltonian
\begin{equation}
\hat{H}=\hbar\omega_\text{m}\hat{b}^\dagger\hat{b}+\sum_{i=1,2}\hbar\bigl[\Delta_i\hat{a}^\dagger_i\hat{a}_i+g_i\hat{a}^\dagger_i\hat{a}_i\bigl(\hat{b}+\hat{b}^\dagger\bigr)\bigr]
+\hbar J\bigl(\hat{a}_1\hat{a}_2^\dagger+\hat{a}_1^\dagger\hat{a}_2\bigr)+\sum_{i=1,2}\hbar\mathcal{E}_i\big(\hat{a}_i+\hat{a}_i^\dagger\bigr).
\end{equation}

Current realisations of such optomechanical systems have a coupling strength $g_i$ that is rather small\cite{Aspelmeyer2014}. This is overcome by means of strong classical driving, which allows us to approximate $\hat{H}$ by means of a Hamiltonian that is quadratic in the operators, and which therefore leads to linear equations of motion. This process is detailed elsewhere in the literature\cite{Aspelmeyer2014}, so we will only list the key steps to linearisation. First, we start from the master equation governing the full system
\begin{equation}
\tfrac{\rmd}{\rmd t}\rho=-\tfrac{\imath}{\hbar}\bigl[\hat{H},\rho\bigr]+\sum_{i=1,2}\mathcal{D}_{\bar{N}_i,\kappa_i,\hat{a}_i}[\rho]+\mathcal{D}_{\bar{N}_\text{m},\gamma_\text{m},\hat{b}}[\rho],
\end{equation}
defining $\bar{N}_\text{m}$ as the average occupancy of the mechanical bath. Next, rewrite
\begin{align}
\hat{a}_i&\to\hat{a}_i+\alpha_i,\ \text{and}\\
\hat{b}&\to\hat{b}+\beta,
\end{align}
where the $\alpha_i$ and $\beta$ are complex numbers whose values will be determined self-consistently. The terms in the resulting master equation can be sorted by their order, i.e., constants, or linear, quadratic, or cubic in the field operators. Constants can be ignored, since they do not affect the dynamics. The linear terms can be eliminated by solving a set of equations that define the $\alpha_i$ and $\beta$ in terms of each other and of the $\mathcal{E}_i$. Operating under the assumption that $\lvert\mathcal{E}_i\rvert$ is large enough so that $\lvert\alpha_i\rvert\gg1$, we can ignore the cubic terms. Finally, defining
\begin{equation}
G_i:=g_i\alpha_i,
\end{equation}
we obtain the so-called linearised optomechanical Hamiltonian
\begin{equation}
\hat{H}_\text{lin}=\hbar\omega_\text{m}\hat{b}^\dagger\hat{b}+\sum_{i=1,2}\hbar\bigl[\Delta_i\hat{a}^\dagger_i\hat{a}_i+\bigl(G_i^\star\hat{a}_i+G_i\hat{a}_i^\dagger\bigr)\bigl(\hat{b}+\hat{b}^\dagger\bigr)\bigr]
+\hbar J\bigl(\hat{a}_1\hat{a}_2^\dagger+\hat{a}_1^\dagger\hat{a}_2\bigr),
\end{equation}
with the rest of the master equation unchanged:
\begin{equation}
\tfrac{\rmd}{\rmd t}\rho=-\tfrac{\imath}{\hbar}\bigl[\hat{H}_\text{lin},\rho\bigr]+\sum_{i=1,2}\mathcal{D}_{\bar{N}_i,\kappa_i,\hat{a}_i}[\rho]+\mathcal{D}_{\bar{N}_\text{m},\gamma_\text{m},\hat{b}}[\rho].
\end{equation}
The equations of motion for these new operators read
\begin{subequations}
\begin{align}
\tfrac{\rmd}{\rmd t}\hat{a}_1&=-\bigl(\imath\Delta_1+\tfrac{\kappa_1}{2}\bigr)\hat{a}_1-\imath J\hat{a}_2-\imath G_1\bigl(\hat{b}+\hat{b}^\dagger)+\sqrt{\kappa_1}\hat{a}_{\text{in},1},\\
\tfrac{\rmd}{\rmd t}\hat{a}_2&=-\bigl(\imath\Delta_2+\tfrac{\kappa_2}{2}\bigr)\hat{a}_2-\imath J\hat{a}_1-\imath G_2\bigl(\hat{b}+\hat{b}^\dagger)+\sqrt{\kappa_2}\hat{a}_{\text{in},2},\ \text{and}\\
\tfrac{\rmd}{\rmd t}\hat{b}&=-\bigl(\imath\omega_\text{m}+\tfrac{\gamma_\text{m}}{2}\bigr)\hat{b}-\imath\bigl(G_1^\ast\hat{a}_1+G_2^\ast\hat{a}_2+G_1\hat{a}_1^\dagger+G_2\hat{a}_2^\dagger\bigr)+\sqrt{\gamma_\text{m}}b_{\text{in},\text{m}},
\end{align}
\label{eq:OMEoM}
\end{subequations}
where we have defined the input field operators similarly to the previous section. Specifically, for $i,j=1,2$, we have
\begin{align}
\langle\hat{a}_{\text{in},i}(t)\rangle&=0,\\
\langle\hat{a}_{\text{in},i}^\dagger(t)\hat{a}_{\text{in},j}(t^\prime)\rangle&=\bar{N}_i\delta_{i,j}\delta(t-t^\prime),\ \text{and}\\
\langle\hat{a}_{\text{in},i}(t)\hat{a}_{\text{in},j}^\dagger(t^\prime)\rangle&=\bigl(\bar{N}_i+1)\delta_{i,j}\delta(t-t^\prime),
\end{align}
as well as
\begin{align}
\langle\hat{b}_{\text{in},\text{m}}(t)\rangle&=0,\\
\langle\hat{b}_{\text{in},\text{m}}^\dagger(t)\hat{b}_{\text{in},\text{m}}(t^\prime)\rangle&=\bar{N}_\text{m}\delta(t-t^\prime),\\
\langle\hat{b}_{\text{in},\text{m}}(t)\hat{b}_{\text{in},\text{m}}^\dagger(t^\prime)\rangle&=\bigl(\bar{N}_\text{m}+1)\delta(t-t^\prime),\ \text{and}\\
\langle\hat{b}_{\text{in},\text{m}}(t)\hat{a}_{\text{in},i}^\dagger(t^\prime)\rangle&=\langle\hat{b}_{\text{in},\text{m}}^\dagger(t)\hat{a}_{\text{in},i}(t^\prime)\rangle=0.
\end{align}
Since we have a concrete model in mind of three bosonic modes (two electromagnetic and one mechanical) we can relate the average occupancy of the baths to their temperatures, by means of the formulae ($i=1,2,\text{m}$)
\begin{equation}
\bar{N}_i=\frac{1}{\exp\bigl[\hbar\omega_i/(k_\text{B}T_i)\bigr]-1},
\end{equation}
with $T_i$ being the absolute temperature of the bath.

To make the connection with our previous formalism, we first rewrite equations~(\ref{eq:OMEoM}) in the frequency domain:
\begin{align}
-\imath\omega\hat{a}_1&=-\bigl(\imath\Delta_1+\tfrac{\kappa_1}{2}\bigr)\hat{a}_1-\imath J\hat{a}_2-\imath G_1\bigl(\hat{b}+\hat{b}^\dagger)+\sqrt{\kappa_1}\hat{a}_{\text{in},1},\\
-\imath\omega\hat{a}_2&=-\bigl(\imath\Delta_2+\tfrac{\kappa_2}{2}\bigr)\hat{a}_2-\imath J\hat{a}_1-\imath G_2\bigl(\hat{b}+\hat{b}^\dagger)+\sqrt{\kappa_2}\hat{a}_{\text{in},2},\ \text{and}\\
-\imath\omega\hat{b}&=-\bigl(\imath\omega_\text{m}+\tfrac{\gamma_\text{m}}{2}\bigr)\hat{b}-\imath\bigl(G_1^\ast\hat{a}_1+G_2^\ast\hat{a}_2+G_1\hat{a}_1^\dagger+G_2\hat{a}_2^\dagger\bigr)+\sqrt{\gamma_\text{m}}b_{\text{in},\text{m}}.
\end{align}
Next, we solve the last of these equations for $\hat{b}$. We take $\Delta_i\approx\omega_\text{m}$ and assume operation in the sideband-resolved regime ($\omega_\text{m}\gg\kappa_i$), which together allow us to eliminate contributions from creation operators in the equation for $\hat{b}$. Finally, we substitute this solution into the equations for $\hat{a}_i$ ($i=1,2$). In vector form, we find
\begin{multline}
-\imath\omega\begin{pmatrix}
    \hat{a}_1 \\
    \hat{a}_2
  \end{pmatrix}=\begin{bmatrix}
    -\imath\Delta_1-\frac{\kappa_1}{2}-\lvert G_1\rvert^2\chi_\text{m}(\omega) & -\imath J-\chi_\text{m}(\omega)G_1G_2^\ast \\
    -\imath J-\chi_\text{m}(\omega)G_1^\ast G_2&-\imath\Delta_2-\frac{\kappa_2}{2}-\lvert G_2\rvert^2\chi_\text{m}(\omega)
  \end{bmatrix}\begin{pmatrix}
    \hat{a}_1 \\
    \hat{a}_2
  \end{pmatrix}\\
+\begin{pmatrix}
    \sqrt{\kappa_1}\hat{a}_{\text{in},1} \\
    \sqrt{\kappa_2}\hat{a}_{\text{in},2}
  \end{pmatrix}+\begin{pmatrix}
    -\imath G_1\sqrt{\gamma_\text{m}}\chi_\text{m}(\omega) \\
    -\imath G_2\sqrt{\gamma_\text{m}}\chi_\text{m}(\omega)
  \end{pmatrix}\hat{b}_{\text{in},\text{m}},
\end{multline}
where we have defined the mechanical susceptibility
\begin{equation}
\chi_\text{m}(\omega):=\frac{1}{\gamma_\text{m}/2-\imath(\omega-\omega_\text{m})}.
\end{equation}
In preparation for the forthcoming equivalence, we perform a gauge transformation $\hat{b}_{\text{in},\text{m}}\to\imath e^{-\imath\nu}\hat{b}_{\text{in},\text{m}}$, where
\begin{equation}
\nu:=\arg\{\chi_\text{m}(\Omega)\},
\end{equation}
where $\Omega$ is some frequency of interest. We assume that $G_1$ is real, which can always be performed by an appropriate choice of phase, and set $G_2\to G_2e^{\imath\phi}$, where the transformed $G_2$ is also real. For convenience, we also define $\tilde{\chi}_\text{m}(\omega):=e^{-\imath\nu}\chi_\text{m}(\omega)$. We thus obtain, quite simply,
\begin{multline}
-\imath\omega\begin{pmatrix}
    \hat{a}_1\\
    \hat{a}_2
  \end{pmatrix}=\begin{bmatrix}
    -\imath\Delta_1-\frac{\kappa_1}{2}-G_1^2\chi_\text{m}(\omega) & -\imath J-\chi_\text{m}(\omega)G_1G_2e^{-\imath\phi}\\
    -\imath J-\chi_\text{m}(\omega)G_1G_2e^{\imath\phi}&-\imath\Delta_2-\frac{\kappa_2}{2}-G_2^2\chi_\text{m}(\omega)
  \end{bmatrix}\begin{pmatrix}
    \hat{a}_1\\
    \hat{a}_2
  \end{pmatrix}\\
+\begin{pmatrix}
    \sqrt{\kappa_1}\hat{a}_{\text{in},1}\\
    \sqrt{\kappa_2}\hat{a}_{\text{in},2}
  \end{pmatrix}+\begin{pmatrix}
    G_1\sqrt{\gamma_\text{m}}\tilde{\chi}_\text{m}(\omega)\\
    G_2\sqrt{\gamma_\text{m}}\tilde{\chi}_\text{m}(\omega)
  \end{pmatrix}\hat{b}_{\text{in},\text{m}}.
\end{multline}
Before continuing, we note that the off-diagonal elements of the matrix in the first term on the right-hand side of this equation are neither complex conjugates of, nor equal to, each other. It is for this reason that this optomechanical platform gives rise to non-reciprocal behaviour. In the next section we will explicitly show how this platform realises the cascaded system described earlier.

\section{EQUIVALENCE BETWEEN THE TWO SCENARIOS}
Suppose that we are concerned with noise in a bandwidth that is small compared to $\gamma_\text{m}$ but large compared to $\kappa_i$. This situation can be realised by having the mechanical oscillator interact with a third eletromagnetic mode to increase its damping rate\cite{Bernier2017}. Under these circumstances, it is fair to assume that the mechanical susceptibility is no longer a function of frequency. By formally taking $\gamma_\text{m}\gg\lvert\Omega-\omega_\text{m}\rvert$, the equivalence between the optomechanical platform and the cascaded system is exact. This can be seen by comparing the two:
\begin{center}
\begin{tabular}{|c||c|}
\hline
\textbf{Cascaded system} & \textbf{Optomechanical platform}\\
\hline
\hline
$\hat{c}_i$ & $\hat{a}_i$\\
$\hat{c}_{\text{in},i}$ & $\hat{a}_{\text{in},i}$\\
$\hat{c}_{\text{in},3}$ & $\hat{b}_\text{in,m}$\\
$\omega_i$ & $\Delta_i$\\
$\gamma_i$ & $\frac{4G_i^2}{\gamma_\text{m}}$\\
$F$ & $J-\frac{2\imath G_1G_2e^{-\imath\phi}}{\gamma_\text{m}}$\\
\hline
\end{tabular}
\end{center}
Any results derived from the cascaded system formalism, therefore, apply identically to the optomechanical platform. For example, if we set $\phi=\pi/2$, such that $e^{\imath\phi}=\imath$, and $J=2G_1G_2/\gamma_\text{m}$, then we recover $F=0$ and perfect non-reciprocity.

\section{RESULTS}
\begin{figure}[ht]
 \begin{center}
   \includegraphics[scale=0.7]{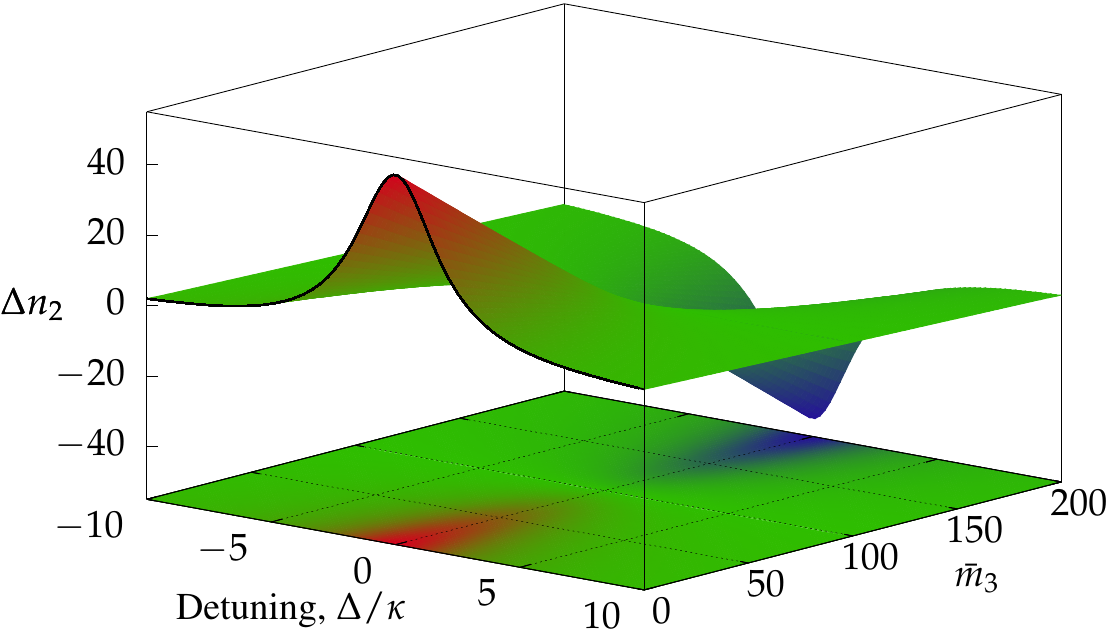}\quad
   \includegraphics[scale=0.7]{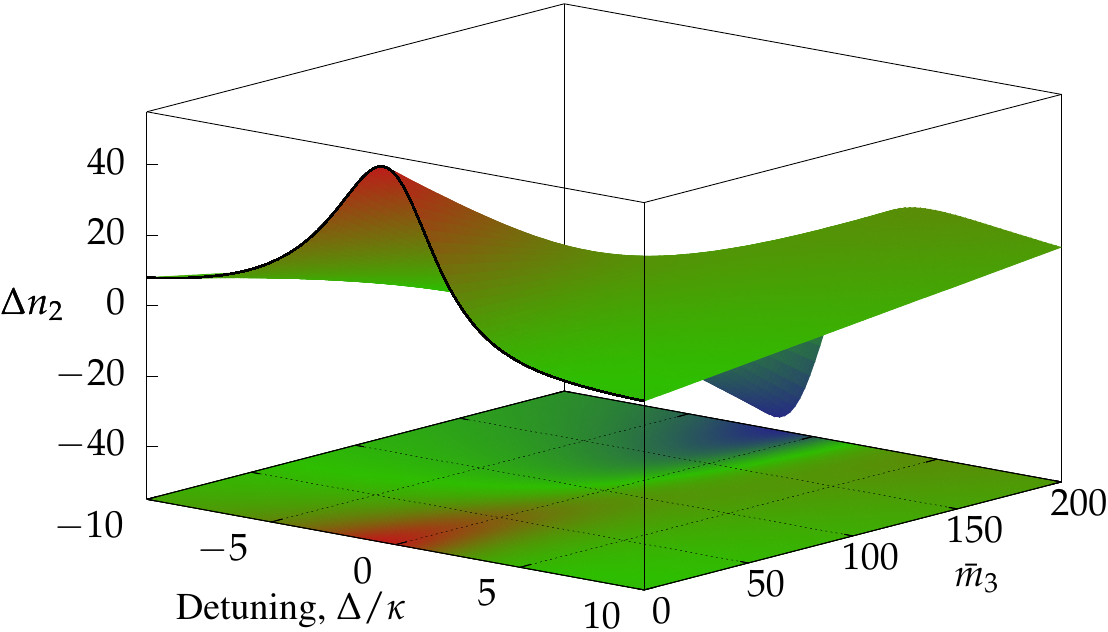}
 \end{center}
 \caption{\label{fig:Deltan2}Change in occupation number of system 2, $\Delta n_2$, as a function of the detuning $\Delta$ between the two modes and the occupancy of the common bath, $\bar{m}_2$. Regions where the thermal noise in the system 2 increases are shown in red, whereas regions where it decreases are shown in blue. We used $\phi=0$, $\bar{m}_1=100$, and $\bar{m}_2=50$. (Left) $F=0$, which means that the thermal noise in system 1 is unaffected. (Right) $F=\kappa$, which means that the system is not fully non-reciprocal; in this case, the control over the thermal noise in system 2 is still present.}
\end{figure}

We are finally in a position to demonstrate how this system leads to a modified flow of thermal noise. To avoid defining heat and flow of heat in the quantum regime, we will base our discussion on the average thermal occupancy of the two field modes, which we write as $\bar{n}_i:=\langle\hat{c}_i^\dagger\hat{c}_i\rangle$ ($i=1,2$) in the notation of the first section. Because of the form of the equations of motion, we know that the state of the field modes will be a thermal state, and so fully characterised by $\bar{n}_i$.

Our intention is to compare two situations, with and without the non-reciprocal link. If we simply removed the coupling between the two systems and the common bath, we would obtain two disconnected field modes, but we would have modified the physics: In the cascaded scenario, each field mode is coupled to two baths, but in this hypothetical case each would be coupled to only a single bath. The way forward is to compare $\bar{n}_i$ with the equivalent quantity, which we denote $\bar{m}_i$, obtained in the formal limit $\lvert\Delta\rvert\to\infty$, where $\Delta:=\omega_1-\omega_2$, whilst keeping $F$, $\kappa_i$, and $\bar{N}_i$ fixed. We use this to define
\begin{equation}
\Delta n_i:=\bar{n}_i-\bar{m}_i.
\end{equation}
The physical interpretation of this quantity is straightforward. A positive (negative) $\Delta n_i$ means an increase (decrease) in thermal noise, brought about by the non-reciprocal link. In the simplest case when $\kappa_1=\kappa_2=\gamma_1=\gamma_2=:\kappa$ and $F=0$, for example,
\begin{align}
\Delta n_1&=0,\ \text{and}\\
\Delta n_2&=\frac{2\kappa^2}{4\kappa^2+\Delta^2}\bigl(\bar{m}_1-\bar{m}_3\bigr).
\end{align}
Under these circumstances, moreover, we have
\begin{align}
\bar{m}_i&=\tfrac{1}{2}\bigl(\bar{N}_i+\bar{N}_3\bigr),\ \text{for}\ i=1,2\ \text{and}\\
\bar{m}_3&=\bar{N}_3.
\end{align}
By way of example, we show in Fig.~\ref{fig:Deltan2} the change in occupation number $\Delta n_2$ of the second mode as a function of the detuning $\Delta$ and the occupancy $\bar{m}_3$ of the common bath. We note that the temperature of the common bath acts as a knob through which the thermal noise of the second mode can be increased or decreased. For the same parameters, when $F=0$ we find that $\Delta n_1=0$ throughout, demonstrating the power of our system to route thermal noise to or away from the second mode without affecting the first.

\section{CONCLUSIONS}
We have presented an optomechanical platform on which we can demonstrate the ability to controllably route thermal noise into or out of an electromagnetic field mode. To analyse this system we constructed a simplified model based on the cascaded systems formalism. Our results show that it is possible to use a heat bath as a knob with which to route thermal noise towards or away from particular systems in a network of quantum devices.

\acknowledgments
We acknowledge funding from the European Union's Horizon 2020 research and innovation program under grant agreement no.\ 732894 (FETPRO HOT). S.B.\ acknowledges support under the Marie Sk\l{}odowska-Curie Actions programme, grant agreement no.\ 707438 (MSCA-IF-EF-ST SUPEREOM).

\end{document}